# Do conspicuous manuscripts experience shorter time in the duration of peer review?

Guangyao Zhang[1], Furong Shang[2], Weixi Xie[1], Yuhan Guo[1], Chunlin Jiang[1], Xianwen Wang[1]*

1. WISE Lab, Institute of Science of Science and S&T Management, Dalian University of Technology, Dalian, China
2. College of geography and environment, Shandong Normal University, Jinan, China

* Corresponding author, xianwenwang@dlut.edu.cn

**Abstract:** A question often asked by authors is how long would it take for the peer review process. Peer review duration has been concerned much by authors and attracted much attention in academia these years. Existing research on this field focuses primarily on a single quantitative dimension. Seldom studies considered that peer review duration is closely related to the attractiveness of manuscripts. This study aims to fill this research gap employing attention economy theory. By analyzing the peer review history from *the British Medical Journal (BMJ)*, we find that a significant negative relationship exists between the peer review duration and altmetric attention score (AAs). Overall, our study contributes to understanding peer review behavior from a new perspective and bridging the divide between peer reviews and altmetrics.

## Introduction

Peer review is a fundamental part of scholarly publishing, which is designed to assess the validity, quality, significance and often the originality of research for publication. Its purposes include maintaining the integrity of science by filtering out invalid or inferior quality articles, and improving the quality of manuscripts that are deemed suitable for publication (Kelly et al., 2014).

Scientists already experiencing too much stress at work has become a common phenomenon in academia (Wang et al., 2012, 2013; Cabanac & Hartley, 2013; Barnett et al., 2019). Similarly, peer review also needs reviewers take time and effort. Although scientists can benefit from peer review practices. For instance, they obtained their reputation via reviews (Righi & Takacs, 2017), gratitude from editors (Bernstein, 2013) and appointments to the journal's editorial board (Tite & Schroter, 2007; Gasparyan et al., 2015). Nowadays, "the growth in scientific production may threaten the capacity for the scientific community to handle the ever-increasing demand for peer review of scientific publications" (Kovanis et al., 2016). This situation has continuously aroused wide attention due to the reviewer crisis in recent years. A number of studies on reviewers' working conditions revealed that reviewing manuscripts has become a serious burden for scientists (Cabanac & Hartley, 2013; Campos-Arceiz et al., 2013; Ramos et al., 2017). Peer review increase scientists' total working hours, and even erodes their leisure time (Wang et al., 2012, 2013; Cabanac & Hartley, 2013; Barnett et al., 2019).

Some studies suggested that it is possible to review a manuscript in less than a day, or even five hours (Ware & Mabe, 2015). However, according to *the Publons' Global State of Peer Review Report* (Publons, 2018b), it takes reviewers a median of 16.4 days to complete a review. Considering the specific work of reviewers, for instance, reviewers often receive multiple invitations to review manuscripts from different journals at the same time, which is a common practice in academia. As a scientist, reviewers usually have heavy research tasks. Therefore, reviewers may need to trade off their limited available time when they should give "5 hours" to review a manuscript (Mrowinski et al., 2016; Huisman & Smits, 2017).

In this article, we attempt to examine a possibility that the peer review duration is closely related to reviewers' interests in a manuscript. We call these interests the attractiveness of a manuscript and measure the attractiveness by altmetric attention score (AAs) because altmetrics reflects people's online attention or general interest, even public engagement (Delli et al., 2017; Starbuck & Purtee, 2017; Hassona et al., 2019). (More detailed rationale and interpretations of the measuring are provided in the Methods section). Specifically, the more conspicuousness, the shorter peer review duration. Using the peer review history of *the British Medical Journal (BMJ)* as a test bed, we explore this question by identifying the relationship between peer review duration and altmetric attention score. This article is organized as follows. In our next section, we briefly review extant researches related to peer review duration. We construct the relationship between peer review duration and attractiveness of manuscripts by citing the attention economy theory. Then follows a description of data sources and methods. We examine a set of peer review history from *BMJ* to analyze the peer review duration and apply regression modeling to test for the influence of peer review duration on AAs. Finally this article is concluded with a discussion of tentative explanations and limitations (Tang et al., 2015).

**Background**

*Practical background*

*peer review duration*

Peer review duration reflects reviewers' behaviors to a certain extent. Scientists always hope to publish their discoveries faster in reputable journals than their competitors. So, they concern much about the peer review duration. The existing studies related to the peer review duration mainly focus mainly on the difference of peer review duration under different conditions. For instance, at the journal level, higher impact factor journals tend to be significantly quicker in moving from submission to acceptance (Pautasso & Schäfer, 2010). At the disciplinary level, differences are apparent between scientific fields. Natural sciences experience a shorter duration than humanitarianism and social sciences (Huisman & Smits, 2017). The editing process also affects peer review duration. Peer review duration is similar for all classes of reviewers, but the completion rate of reviewers known personally by the editor is higher (Mrowinski et al., 2016). In addition, duration is also related to review rounds (Huisman & Smits, 2017). Out of concern for review time, some studies have specifically provided work to automatically extract the peer review duration data and help researchers find the average response times of journals (Bilalli et al., 2020). Overall, existing researches on peer review duration focus primarily on a single quantitative dimension. Seldom do researchers pay attention to the significance of this index from the perspective of scientometrics.

### *Peer review and influences of bibliometrics*

Numerous studies have dived into peer review and influences of bibliometrics, although these studies mainly focused on citations impact. Several studies have investigated the fate of manuscripts that were rejected from journal. The results suggested that most of the initially rejected manuscripts were eventually published in journals with an impact factor lower than that of journals of initial submission (Zoccali et al., 2015; Casnici, Grimaldo, Gilbert, Dondio, et al., 2017), with a significantly less citation (Docherty & Klein, 2017). Therefore, some researchers suggested that peer review plays a part in quality control. However, Siler also pointed that sometimes peer review can lead to excellent papers being rejected (Siler et al., 2015). Researchers can analyze the citations of papers to see whether certain groups of applicants or authors are favored or at a disadvantage (Bornmann, 2011; Card et al., 2020; Zhu, 2021). Other latest studies have discussed the relationship between review progress and citations. Sikdar found that papers receiving fewer rounds of review tended to receive higher numbers of citations once accepted (Sikdar et al., 2017). However, this result was not confirmed in Wolfram's study (Wolfram et al., 2021). Shideler even showed that reviewers' interest in a manuscript may predict its future citation potential (Shideler & Araújo, 2017). It is worth mentioning that, extrapolating from the available literatures, we found that the perspective of altmetrics has received little academic attention.

### *Theoretical background*

#### *Attention economy theory*

We hope to bridge the divide between peer review duration and attractiveness of manuscripts through attention economy theory. Attention economy theory considers the cost and benefit of searching for useful information (Beck & Davenport, 2001). The main cost paid by authors who consume information is their attention on the information age. People's attention becomes more precious because the amount of information is increasing rapidly while people's attention becomes limited. The same is true for scientists: the amount of information faced by modern scientists is far beyond their cognitive abilities. So, how to manage and protect their attention is particularly important for scientists. On the one hand, scientists get as much attention from other scientists as possible, on the other hand, they spend as little attention as possible on screening information.

Some studies use this theoretical perspective to analyze behavior on social media platforms. For instance, after analyzing actions of more than 3 million twitter authors, a study point out that the social media environment is like an attention market, where people produce information to attract attention and contribute attention when consuming information (Rui & Whinston, 2012). Attention economy theory can also be used to evaluate novelty and popularity of authors in social networks (Huberman, 2013). The attention level of the audience depends on the total amount of attention and the total amount of signals they face (Falkinger, 2007).

Other empirical studies suggest that reviewers have some cognitive processing capacity, although this capacity is limited (Roetzel, 2019; Garcia et al., 2019). In addition, they could consider using these sacrificed time for their own research activities (Bernstein, 2013). This fact indicates that scientists can choose levels of efforts for reviews and manuscripts has also been confirmed (Squazzoni et al., 2013; Bianchi et al., 2018). For instance, the results by Serra-Garcia and Gneezy show that when the paper is more interesting, reviewers may adopt lower standards regarding its reproducibility (Serra-Garcia & Gneezy, 2021). Therefore, for reviewers, although reviewing a manuscript may take a day, it is different to decide when to devote their attention to a manuscript (Cabezas Del Fierro et al., 2018). Attention economy theory may explain this behavior to a certain extent: manuscripts may compete for reviewers' attention. That is to say, conspicuous research may be reviewed more quickly. Actually, according to Publons' report, reviewers spending less time on manuscripts that are very poor or, conversely, very good (Publons, 2018a).

**Methods**

*Data*

We select *the British Medical Journal* (*BMJ*) as our data source because *BMJ* provides unique detailed peer review records (Zhang et al., 2022). Research papers submitted to *BMJ* after September 2014 usually have their prepublication history posted on the bmj.com after being accepted (Groves & Loder, 2014). Additionally, our supplementary data comes from Web of Science and altmetric.com.

In this paper, we attempt to use altmetric attention score to reflect the attractiveness of research. Based on social media data of scholarly articles, altmetrics, which are defined as alternatives or complements to traditional metrics, were designed and proposed (Priem & Hemminger, 2010). Several studies believe that altmetrics could reflect the social impact of scientific results to some extent (Bornmann, 2014a, 2014b, 2015). Increasing studies suggest altmetrics data could be useful as an aid to assessing impact (Barnes, 2015; Wang, Fang, Li, et al., 2016; Wooldridge & King, 2019). Proponents of altmetrics approaches have pointed out that new media allow for new avenues of scientific impact assessment. Altmetrics provide a promising approach to complementing scientific impact assessment (Hoffmann et al., 2016; Wang, Fang, & Guo, 2016).

Although measuring the societal impact of researches is difficult for multiple reasons (Thelwall, 2020). Actually, the role of altmetrics in this field has also begun to be questioned (Tahamtan & Bornmann, 2020). However, a number of studies still believe that altmetrics can measure attention (Konkiel et al., 2016; Moed, 2017; Bornmann et al., 2019), popularity (Xia et al., 2016) or public engagement (Khazragui & Hudson, 2015). Specifically, Brigham hold that altmetrics measures the impact of each article through the attention attracted online (Brigham, 2014). Wei believe that altmetric attention score may be used to assess "trending" articles, which are those that are most interesting to and shared by the general public (Wei et al., 2021). Araujo consider that altmetric attention score reveals the instantaneous attention attracted online for articles in news outlets, comments on blogs, number of tweets, and mentions on social media (Araujo et al., 2021). In terms of empirical studies, the results by Andersen and Haustein suggest that tweets reflect the attractiveness of papers for a broader audience. (Andersen & Haustein, 2015). Zhou and Hassona also demonstrated that social media audience size is associated with the popularity of academic articles across multiple web platforms (Zhou et al., 2018; Hassona et al., 2019). Publishers have taken notice of this feature of altmetrics. For instance, the *JAMA* Network has embed altmetric attention score into online articles to help readers identify papers which were recognized as interesting works (2021 American Medical Association, 2021).

Based on previous literature, we call this feature that attracts people's attention or raises readers' interests the attractiveness of manuscripts and measure this feature by the altmetric attention score of each manuscript. So, we believe that altmetric attention score is a suitable variable to reflect the attractiveness of a manuscript.

As to peer review duration, we primarily focus on first decision because duration of the first review round is probably the most important review for authors (Azar, 2007; Casnici, Grimaldo, Gilbert, & Squazzoni, 2017). Some studies, even if they are at the forefront (Xu et al., 2021), hold that publication delay for an article is defined as the number of days from submission date to acceptance date. However, this definition may be unreliable in measuring peer review duration. We take the gap between submitting an original manuscript and receiving a first decision as peer review duration. We corrected the incorrect date in the data processing process. After that，we collected data of 691 articles published between March 2015 and April 2020 from bmj.com. It is worth mentioning that the relationship between the duration of peer reviews and the attractiveness of manuscripts is only one aspect of the discussion. Because the behavior of peer review is very complex and involves many factors. Unfortunately, up to now, only the peer review history provided by bmj.com has enabled us to discuss this question more accurately. So, peer review duration provided by bmj.com is the best choice for empirical analysis in our research. Although our data is limited to one journal, the connection between peer review duration and altmetric attention score discussed in this paper has not been investigated by relevant researches before. Therefore, our research is innovative to some extent, and relevant findings will form the basis for further research in the future. In addition, it should be pointed out that this paper analyzes basic phenomena, which have certain stability over a short period. Therefore, conclusions of this paper have some value for today's researches. At the end of the empirical analysis, we also used the publication history of *Research Policy* (*RP*)，*Journal of Informetrics* (*JoI*) and *Journal of the Association for Information Science and Technology* (*JASIST*) from ScienceDirect.com and Wiley Online Library. We hope to further discuss and expand on conclusions of our research.

## Analysis of Results

### *Basic Descriptives*

Figure 1 shows the distribution of peer review duration in different rounds and different numbers of reviewers. As shown in Figure 1. a), line in orange connects the median of peer review duration in each round (the first five rounds). The peer review duration generally shows a decreasing trend with the increase in number of review rounds. Similarly, we also describe the distribution of revised duration in different rounds in Figure 1. b), and get similar results. With the deepening of the review process, problems in manuscripts will be clarified and the interaction duration between journals and authors will be shortened. We observed the distribution of the peer review duration with the number of reviewers less than 6 (this part of manuscripts accounts for 93.59%). In different review rounds, peer review duration is also related to the number of reviewers. Generally speaking, the more reviewers there are, the longer the duration it takes. Because peer review duration of a manuscript is determined by the slowest reviewer. The more reviewers, the more likely it is to encounter a long duration.

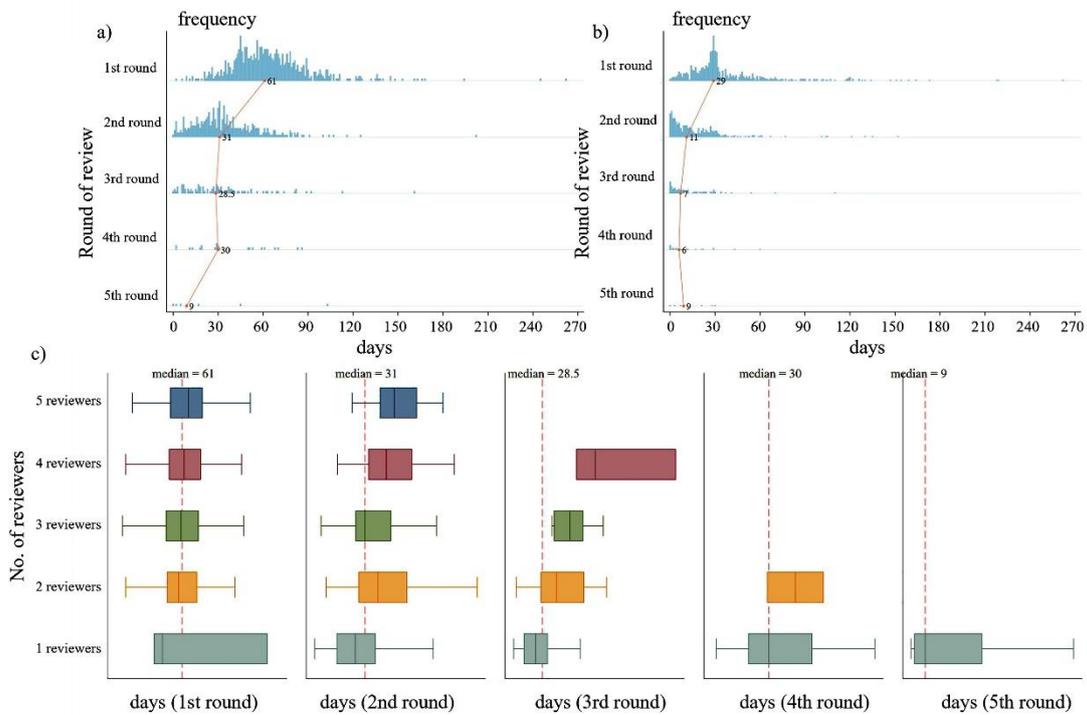

**FIGURE 1**

**a)** peer review duration in different rounds. **b)** revised duration in different rounds. **c)** peer review duration under different number of reviewers (first decision).

In Figure 2, we visualized the relationship between peer review duration and altmetric attention score using scatterplots. We constructed kernel-weighted local polynomial smoothing scatterplot (Li & Agha, 2015) to enhance the smoothed lines. In addition, to overcome overcrowding in scatterplots, we constructed binned scatterplots (Chetty et al., 2014). The x-axis variable was divided into an equal number of groups, the mean values of the x- and y-axis variables in these bins were calculated, and then a binned scatterplot (with a trend line) was plotted, so that the relationship between the variables could be clearly visualized.

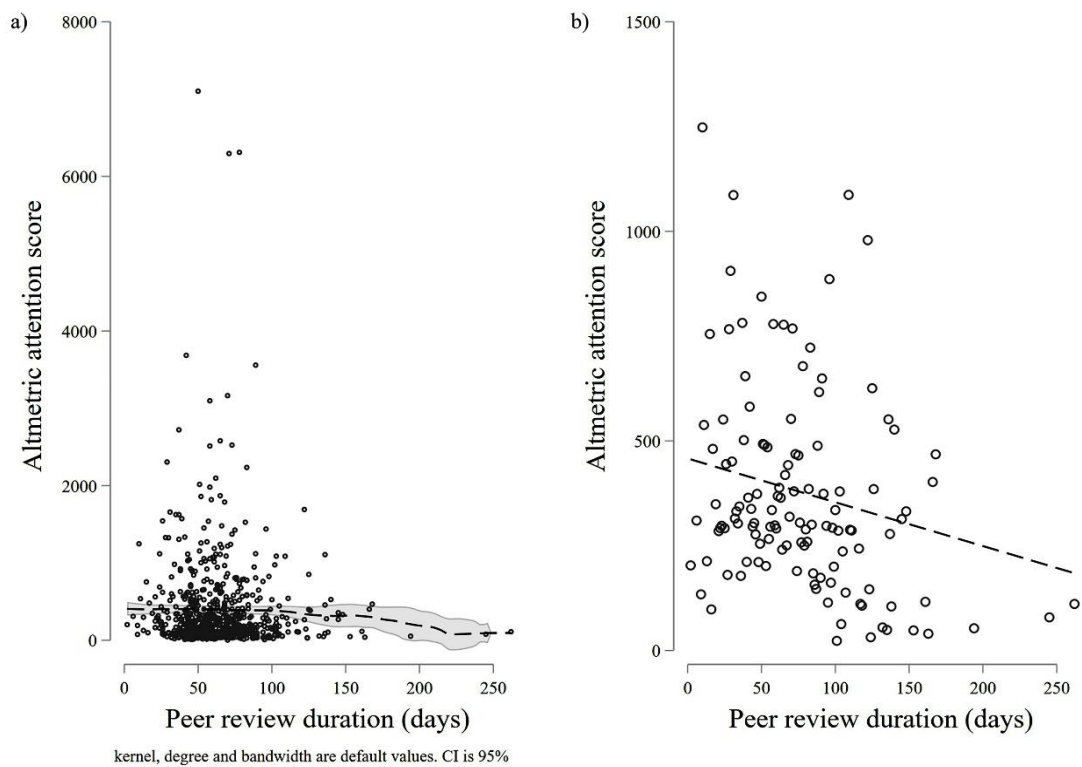

**FIGURE 2**
a) Kernel-weighted local polynomial smoothing scatterplot of peer review duration (days) and altmetric attention score. b) Binned scatterplot of peer review duration (days) and altmetric attention score.

## *Statistical Regression*

The descriptive statistics suggest that there is a negative relationship between peer review duration and altmetric attention score. We next test whether this relationship hold when controlling for confounding factors such as authors, reviewers. Our null hypothesis is straightforward:

**H0:** Manuscripts with short peer review duration mean more conspicuous.

The alternative hypothesis is:

**H1:** There is no statistical difference in the relationship between peer review duration and attractiveness of manuscripts.

The regression model is as follows:

$Y_i = X_i + \varepsilon_i$ where Y represents the attractiveness of manuscripts, $X_i$ is a vector of characteristics impacting $Y_i$, and $\varepsilon_i$ is an error term. Table 1 presents variables and measures in this section.

## *Measurement*

*Dependent v*ariable. We use altmetric attention score to measure the attractiveness of manuscripts.

*Independent variables.* Our explanatory variable is the peer review duration of a manuscript, which is a time gap (days) between submitting an original manuscript and receiving a first decision.

*Control variables.*

On the basis of previous studies (Haustein et al., 2015; Onodera & Yoshikane, 2015; Tahamtan et al., 2016), we take some factors that may be related to the altmetric attention score as control variables, including paper related factors and author(s) related factors. AT, DU, TI, AU, PG, NR, FU and RE were finally determined as control variable. Specific variables and metrics are shown in Table 1. In addition, considering that there may be obvious differences in altmetric attention score of articles in different months, we set monthly dummy variable (PU) to control the time fixed effect. We also set the reviewer dummy variable (RE) to control the reviewer fixed effect. In addition, we consider the OA statue of manuscripts in further exploration.

Considering the accumulation of altmetric attention score, we choose articles published before 2020 which were reviewed by less than 6 people to avoid the influence of the extreme value of the number of reviewers. We also delete the record of only one reviewer (there are only 3 items). Summary descriptive and correlation statistics (Table 2 and 3) indicate no significant issues for the regression.

TABLE 1. Variable description.

| Construct | Variable | Type | Description |
|---|---|---|---|
| Dependent (Attractiveness) | AT | count | Total altmetric attention score counts |
| Independent (Peer review duration) | DU | count | The time gap (days) between submitted original manuscript and received first decision |
|  | DU_E | count | The gap between submitted original article and revised manuscript |
| Control | TI | count | Length of each review |
|  | AU | count | No. of coauthors |
|  | PG | count | No. of pages |
|  | NR | count | No. of references |
|  | FU | dummy | 1 if paper is funded; 0 if no funded |
|  | PU | dummy | Online publication time (year and month) |
|  | RE | count | No. of reviewers |
|  | OA | dummy | 1 if paper is OA; 0 if Non-OA |

TABLE 2. Summary of descriptive statistics (N = 582).

| Variable | M | Median | SD | Min | Max |
|---|---|---|---|---|---|
| AT | 384.42 | 193.50 | 638.85 | 3 | 7102 |
| DU | 64.61 | 61.00 | 27.46 | 9 | 262 |
| TI | 17.63 | 17.00 | 4.94 | 8 | 36 |
| AU | 11.24 | 8.00 | 19.31 | 1 | 294 |
| *PG* | 10.00  | 10.00  | 2.65  | 4 | 27  |
| *NR* | 47.26  | 41.00  | 25.42 | 0 | 268 |
| *FU* | 0.89   | 1.00   | 0.31  | 0 | 1   |
| *RE* | 3.37   | 3.00   | 1.00  | 2 | 5   |

TABLE 3. Correlation matrix.

| Variable | 1 | 2 | 3 | 4 | 5 | 6 | 7 | 8 |
|----------|------|------|-------|-------|------|------|------|------|
| *1 AT*   | 1.00 |      |       |       |      |      |      |      |
| *2 DU*   | -0.06 | 1.00 |      |       |      |      |      |      |
| *3 TI*   | -0.05 | 0.02 | 1.00  |       |      |      |      |      |
| *4 AU*   | -0.04 | 0.02 | 0.062 | 1.00  |      |      |      |      |
| *5 PG*   | 0.01  | 0.08 | 0.14  | 0.09  | 1.00 |      |      |      |
| *6 NR*   | 0.06  | 0.10 | 0.055 | -0.01 | 0.53 | 1.00 |      |      |
| *7 FU*   | -0.05 | 0.04 | 0.021 | 0.07  | 0.01 | 0.07 | 1.00 |      |
| *8 RE*   | 0.02  | 0.08 | -0.085 | 0.01 | 0.05 | 0.02 | 0.05 | 1.00 |

### *Regression Results*

Altmetric attention score is a discrete variable involving non-negative integers, which represents a typical count variable. Therefore, we adopted "*ln(AT)*" and "*ln(DU)*" as the dependent and independent variable instead of "*AT*" and "*DU*". We started this research with OLS and chose STATA Version 15.1 to run the analysis.

As Table 4 illustrates, we only include PU in m1, add RE to m1 as m2, add PU to m1 as m3, add RE and PU to m1 as m4, add other control variables to m1 as m5, and add all variables as m6. It is obvious that a significant negative relationship exists between peer review duration and altmetric attention score. All p values associated with the overall model are less than 0.05. It should be noted that symbols of the control variables in this regression model have changed, but this change does not affect our results. Unlike the effect of explaining main variables that should be paid attention to in this research, the control variables have almost no substantive significance, so we can safely omit them (Liang & Zeger, 1995; Hünermund & Louw, 2020).

TABLE 4. Regression results.

|       | m1<br>Basic | m2<br>RE | m3<br>PU | m4<br>RE & PU | m5<br>Controls | m6<br>All variables |
|-------|-------------|----------|----------|---------------|----------------|---------------------|
| lnDU  | -0.265**    | -0.282*** | -0.252** | -0.263**      | -0.274***      | -0.289**            |
|       | (0.104)     | (0.105)  | (0.112)  | (0.113)       | (0.105)        | (0.114)             |
| TI    |             |          |          |               | -0.022**       | -0.013              |
|       |             |          |          |               | (0.010)        | (0.011)             |
| AU    |             |          |          |               | -0.003         | -0.005**            |
|       |             |          |          |               | (0.002)        | (0.002)             |
| PG    |             |          |          |               | -0.009         | -0.020              |
|       |             |          |          |               | (0.022)        | (0.023)             |
| NR    |             |          |          |               | 0.005**        | 0.007***            |

|        | (1)       | (2)       | (3)       | (4)       | (5)       | (6)       |
|--------|-----------|-----------|-----------|-----------|-----------|-----------|
|        |           |           |           |           | (0.002)   | (0.002)   |
| FU     |           |           |           |           | 0.015     | -0.113    |
|        |           |           |           |           | (0.164)   | (0.164)   |
| _cons  | 6.344***  | 6.335***  | 5.748***  | 5.744***  | 6.662***  | 6.040***  |
|        | (0.430)   | (0.433)   | (0.556)   | (0.555)   | (0.480)   | (0.609)   |
| N      | 582       | 582       | 582       | 582       | 582       | 582       |
| adj. $R^2$ | 0.007 | 0.007     | 0.073     | 0.071     | 0.019     | 0.088     |
| PU     | No        | No        | Yes       | Yes       | No        | Yes       |
| RE     | No        | Yes       | No        | Yes       | No        | Yes       |

Standard errors in parentheses $^* p < 0.1$, $^{**} p < 0.05$, $^{***} p < 0.01$

### Robustness Tests

We further adopted robustness tests. In m7, we try to replace the dependent variable. Since the altmetric attention score changes with time, we want to verify that our conclusion is valid when using the altmetric attention score at a different time point. (Unfortunately, two items went missing from the data collected on June 27, 2020.) In m1 to m6, we used the data from all articles. However, manuscripts with different review rounds may have heterogeneity. In m8, we attempted to change the sample size, and use manuscripts that have only experienced the first review round, to carry out regression analysis. In order to eliminate the influence of a model setting, we want to confirm whether the conclusion is valid when the model changes in m9. We then specified a negative binomial regression (Nbr) to cater for the count data nature of altmetric attention score. In fact, we also conducted Nbr on samples whose altmetric attention score was collected on June 27, 2020 and a subsample that only experienced the first round of peer review, and got similar results. In m10, the right end of the data is winsorized on 97.5th percentile and then regressed for altmetric attention score are prone to right-biased distribution. The results (Table 5.) showed that DU is significant and all p values are less than 0.05.

TABLE 5. Regression results (Robustness Tests).

|       | m7          |          | m8          |          | m9          |          | m10         |          |
|-------|-------------|----------|-------------|----------|-------------|----------|-------------|----------|
|       | Change time |          | First round |          | Nbr         |          | Winsorized  |          |
| lnDU  | -0.287**    | (0.115)  | -0.470***   | (0.172)  |             |          | -0.289**    | (0.112)  |
| DU    |             |          |             |          | -0.004**    | (0.002)  |             |          |
| TI    | -0.013      | (0.011)  | 0.012       | (0.018)  | -0.010      | (0.010)  | -0.013      | (0.010)  |
| AU    | -0.005**    | (0.002)  | -0.002      | (0.002)  | -0.006***   | (0.001)  | -0.004**    | (0.002)  |
| PG    | -0.022      | (0.023)  | -0.044      | (0.039)  | -0.017      | (0.022)  | -0.021      | (0.022)  |
| NR    | 0.006***    | (0.002)  | 0.006       | (0.005)  | 0.007***    | (0.002)  | 0.007***    | (0.002)  |
| FU    | -0.131      | (0.164)  | -0.276      | (0.249)  | -0.011      | (0.155)  | -0.132      | (0.155)  |
| _cons | 6.109***    | (0.630)  | 6.588***    | (0.813)  | 5.301***    | (0.406)  | 6.033***    | (0.603)  |
| N     | 580         |          | 258         |          | 582         |          | 582         |          |
| adj. $R^2$ | 0.091  |          | 0.060       |          | -           |          | 0.090       |          |
| ll    | -842.942    |          | -344.913    |          | -3947.969   |          | -835.390    |          |

| | | | | |
|---|---|---|---|---|
| aic | 1819.885 | 815.825 | 8031.937 | 1804.780 |
| bic | 2112.208 | 1039.662 | 8328.857 | 2097.333 |
| PU | Yes | Yes | Yes | Yes |
| RE | Yes | Yes | Yes | Yes |

Standard errors in parentheses $^{*} p < 0.1$, $^{**} p < 0.05$, $^{***} p < 0.01$

### *Further exploration*

In our research, we used data provided by bmj.com for analysis, which verified our hypothesis. We are wondering that whether we can extend our research findings to a broader context or not. If we replace the peer review duration with the time gap between submitting original manuscripts and receiving the final revision, could we reach the same conclusion? Here we calculated the correlation coefficients between peer review duration (first decision) and peer review duration (final revision). The spearman correlation coefficient of the first round was the highest, which was 0.768. When more rounds were added, the correlation coefficient decreased to 0.555.

Further, we hope that we can extrapolate our conclusions. Elsevier and Wiley are two major academic publishers and involve many disciplines. More importantly, there are received time and revised time on webpages of published papers. Of course, revised time here refers to the last revised (Cabanac & Hartley, 2013). We hope that we can further explore the relationship between peer review duration and altmetric attention score. We selected three journals with high recognition in science of science, which are related to management, information & library science, and computer science: *Research Policy* (2013-2019), *Journal of the Association for Information Science and Technology* (2014-2020) and *Journal of Informetrics* (2013-2019). We downloaded data of three journals from WoS (including two types: article or review) and obtained the altmetric attention score for each paper from altmetric.com. We calculated the peer review duration (final revision) by manuscript revised time minus manuscript received time. We controlled TI, AU, PG, NR, FU, OA and PU in regression. Peer review duration (final revision) and ln(altmetric attention score + 1) presented a significant negative relationship, which was significant at 0.05 level. In fact, Nbr also received similar results.

TABLE 6. Regression results (Further exploration).

| | *JASIST* | | *JoI* | | *RP* | |
|---|---|---|---|---|---|---|
| lnDU_E | -0.137*** | (0.046) | -0.203** | (0.094) | -0.263*** | (0.075) |
| TI | 0.002 | (0.007) | -0.017 | (0.011) | -0.020* | (0.010) |
| AU | 0.024 | (0.018) | 0.115** | (0.053) | 0.119*** | (0.040) |
| PG | -0.022** | (0.009) | -0.032** | (0.013) | 0.011 | (0.015) |
| NR | 0.005*** | (0.001) | 0.007*** | (0.002) | -0.000 | (0.001) |
| FU | -0.029 | (0.063) | -0.224 | (0.159) | 0.152 | (0.098) |
| OA | 0.621*** | (0.082) | 0.411** | (0.187) | 0.531*** | (0.098) |
| cons | 1.293*** | (0.486) | 3.341*** | (0.476) | 1.895*** | (0.591) |
| N | 1114 | | 538 | | 911 | |

| | | | |
|---|---|---|---|
| adj. $R^2$ | 0.138 | 0.138 | 0.083 |
| PU | Yes | Yes | Yes |

**Discussion and Conclusion**

In contrast to previous studies that focused mainly on the relationship between peer review processes and citation impacts, this study contributes to revealing the relationship between peer review duration and the attractiveness of papers. At the beginning of this paper, manuscripts may be competing for the attention of busy reviewers. On this basis, through combing existing researches and attention economy theory, our research puts forward an explanation: there is a negative relationship between peer review duration and the attractiveness of papers.

Our research has following enlightenments. Firstly, in terms of science publishing, it is important to understanding reviewers' behavior, especially their "trade-off" in review progress. As quality control is one of the important functions of peer review, scientists usually devote considerable time and effort for reviewing. Reviewers need to hold objective selection criteria, even in the face of interesting manuscripts. Secondly, some journals, especially some high-quality journals, such as *Nature* and Nature journals, pay attention to the attractiveness of manuscripts primarily. The research should be important for the discipline and (or) interdisciplinary interest and (or) has real world implications (Nature Human Behaviour, 2019). The author must be aware of this point in the manuscripts submitted that the manuscripts' attractiveness may affect reviewers' reviewing speed. Thirdly, the reviewers' work is an "unrequited contribution" to the scientific community. Reviewers are generally not paid for their review work, and that reviews are most anonymous. Few incentives choose to give high priority to this work (Azar, 2007; Moizer, 2009). However, an interesting manuscript that can broaden reviewers' horizons may also be an implicit reward obtained by reviewers after finishing work. Borrowing the words of Engers and Gans (Engers & Gans, 1998), to the wish "to keep up with current ideas and new results". Peer review is also an available path that updates knowledge without incurring financial expenses (Gasparyan et al., 2015). Finally, quantitative social science is entering the era of open science (Zhang et al., 2021). Journals such as *BMJ* and *Nature Communications* have provided abundant peer review data, while large-scale publishing time data provided by publishers such as Elsevier and Wiley have great value in analyzing scientists' behavior. With more information on scientists' behavior, we could focus on much more effective quantitative researches in the field of the sociology of science (D. Lazer et al., 2009; D. M. J. Lazer et al., 2020). This paper is still a preliminary discussion, but there is more content to be further explored in the future.

*Limitations and Future Study*

We acknowledge that our research has several limitations. Firstly, for the sake of historical data, we adopted cross-sectional data, including the peer review history provided by bmj.com and altmetric attention score provided by altmetric.com. Since peer review duration data was formed before altmetric attention score, we think that logically, the former should be an independent variable, while the latter should be a dependent variable. Actually, the direction and significance of these coefficients did not change when we swapped the independent and dependent variable. Although we can find a binary relation between peer review duration and the attractiveness of papers, it needs the support of diachronic data for further causal inference. Secondly, there are complex influencing factors behind the reviewers' behavior, and the review time may vary with the complexity of topics, the potential of manuscripts and the experience of reviewers (Publons, 2018b). The quality and issue period of journals can also affect the time spent of reviewing manuscripts. Because related confounding variables are not measured in the data, this study cannot investigate the potential impact of selectivity error by strict statistical means. If more data is available in the future, this may be a possible direction. Thirdly, we only use data from one journal for empirical analysis, so the basic model found in this study is more instructive for us to understand the relationship between peer review duration and altmetric attention score, and its reliability needs to be tested with a wide range of data. Although we tried to extrapolate our conclusions using the data provided by Elsevier and Wiley, we are still cautious about this approach. Finally, this paper only considers that the gap between submitting original manuscripts and receiving the first decision reflects the peer review duration, and uses altmetric attention score to reflect the attractiveness of research. However, with current review behavior, this feature may have more indicators to reflect. Therefore, future research in this area may need to further subdivide different features.

**Acknowledgments**

The authors would like to thank Yuqi Wang for his advice and support in data analysis and the writing of this paper. We gratefully acknowledge the grant from the National Natural Science Foundation of China (71673038, 71974029).